\documentclass{aa}

\usepackage{graphics}
\usepackage{psfig}

\begin{document}

\title
{Surface photometry of new nearby dwarf galaxies\thanks{Based
on observations obtained with the Apache Point Observatory 3.5-meter
telescope, which is owned and operated by the Astrophysical Research
Consortium.}}

\author{L. N. Makarova\inst{1,2}
	\and
	I. D. Karachentsev\inst{1}
	\and
	E. K. Grebel\inst{3}
	\and
	O. Yu. Barsunova\inst{4}}

\institute{Special Astrophysical Observatory of the Russian Academy of Sciences, Nizhnij Arkhyz 369167,
Karachaevo-Cherkessia, Russia\\
e-mail: {\tt lidia@sao.ru, ikar@luna.sao.ru}
\and
Isaac Newton Institute of Chile, SAO Branch
\and
Max Plank Institute for Astronomy, K\"onigstuhl 17, D-69117 Heidelberg, Germany\\
e-mail: {\tt grebel@mpia-hd.mpg.de}
\and
Astronomical Institute, St. Petersburg State University, 198904 St. Petersburg, Russia\\
e-mail: {\tt monoceros@mail.ru}
}

\abstract{
We present CCD surface photometry of 16 nearby dwarf galaxies, many of
which were only recently discovered.  Our sample comprises both isolated
galaxies and galaxies that are members of nearby galaxy groups.
The observations were obtained in the Johnson B and V bands (and in
some cases in Kron-Cousins I).
We derive surface brightness profiles, total magnitudes, and integrated
colors.  For the 11 galaxies in our sample with distance estimates
the absolute B magnitudes lie in the range of  
$-10 \ga M_B \ga -13$.  The central surface brightness ranges from
22.5 to 27.0 mag arcsec$^{-2}$.
Most of the dwarf galaxies show exponential light
profiles with or without a central light depression. Integrated
radial color gradients, where present, appear to indicate a more centrally
concentrated younger population and a more extended older population.
}

\maketitle

\section{Introduction}

Dwarf irregular (dIrr) and dwarf spheroidal (dSph) galaxies account
for 80--90\% of the total population of galaxies.  While there are
common trends in their global characteristics such as luminosity,
surface brightness, and metallicity, dwarf galaxies may differ 
significantly in details of their star formation histories and 
evolutionary state.  Nearby dwarf galaxies out to about 5 Mpc have
the advantage that we can study both their integrated properties
through medium-sized ground-based telescopes, and their detailed 
stellar content through high-resolution observations with the Hubble
Space Telescope or the new 8m to 10m-class telescopes.  We are
carrying out a large project to study both the integrated properties
and the resolved stellar content of dwarf galaxies in the Local Volume
($V < 500$ km s$^{-1}$),
a necessary precondition for the understanding of
the evolution of unresolved dwarf galaxies at larger distances.  
Hopp \& Schulte-Ladbeck (\cite{hs}), Karachentseva et al.\ 
(\cite{karachentseva96}), Bremnes et al.
(\cite{bremnes98};\cite{bremnes99}), Makarova (\cite{makarova}), and
Jerjen et al. (\cite{jerjen}) presented the results of surface CCD photometry
of many nearby dwarf galaxies within and outside of groups out to 
a distance of 10 Mpc.  Despite this considerable observational
progress, for more than 3/4 of the dwarf galaxies of the Local Volume 
neither surface brightness profiles nor magnitudes and colors have been
measured yet, nor have these galaxies been imaged with modern CCD detectors or
reliably classified by morphological type.

Over the last three years Karachentseva and Karachentsev with their
co-workers have carried out a search for new nearby dwarf galaxies on
the basis of the POSS-II and ESO/SERC sky surveys, covering 97\%
of the sky. Their survey resulted
in detection of about 600 dwarf systems more than half of which were missing
in the catalogues of known galaxies. Subsequent follow-up
observations of these
galaxies in the 21 cm hydrogen radio line (Huchtmeier et al.\ 
\cite{huchtmeier}) confirmed that many of these objects are nearby with
a median radial velocity of $\sim$ 1200 km s$^{-1}$. It should be emphasized
that during the last two decades the total number of probable Local Volume dwarf
galaxies has increased by a factor of two due to new detections
and amounts to $\sim$ 360.

Dwarf galaxy candidates from the latest surveys were included in our
ongoing program of ground-based imaging follow-up observations.  For  
these galaxies we adopt the following naming convention:
``KK'' (Karachentseva \& Karachentsev \cite{kk98}),
``KKSG'' (Karachentsev et al.\ \cite{k2000}), and ``KKH'' (Karachentsev et al.\
\cite{k2001a}), followed by a running number corresponding to the 
line in each of the respective catalogues.  This sample was supplemented
by several low-surface-brightness (LSB) dwarfs located in
the M~81 group detected in other surveys.

\section{Observations}

From the catalogs mentioned in the preceding section we selected
a sample of 35 most likely nearby galaxies in the appropriate coordinate
range. The choice was based on their low radial velocity ($<$ 500 km~s$^{-1}$)
or their visual impression on the photographic plates (size, granulation).
A subset of 16 galaxies were observed successfully.
The observations of these nearby dwarf galaxies were made by I.D.~Karachentsev
and E.K.~Grebel on 2000 February 3 and 9 at the 3.5-m telescope of
Apache Point Observatory (APO) in New Mexico, USA.  We carried out direct 
imaging
using the Seaver Prototype Imaging camera (SPIcam), which is equipped with
a backside-illuminated SITe CCD (2048$\times$2048 pixels, pixel scale
0.14$\arcsec$ pixel$^{-1}$, field of view 4.78$\arcmin \times$ 4.78$\arcmin$).
During the read-out a 2 by 2 binning was applied,
resulting in a pixel scale of 0.28$\arcsec$ pixel$^{-1}$.

The basic parameters of the observed galaxies are listed
in Table~1. The first column contains the galaxy name, the second and third
columns its coordinates for the epoch J2000, the forth and
fifth columns specify the Galactic extinction in the B and V filters
according to Schlegel et al.\ (\cite{schlegel}), the sixth column contains
the angular size of the semimajor axis of the galaxy, the seventh column 
the galaxy's
morphological type, and the eighth gives the radial velocity reduced
to the Local Group centroid.

The observing log is presented in Table~2.  Most of our data were obtained
using the standard Johnson B and V filters for SPIcam.  In a few cases we
used the Kron-Cousins I filter.  Detailed filter data are available from 
the APO web site at {\tt http://www.apo.nmsu.edu/Instruments/filters/ broad.html}.

The seeing ranged from $1''$ 
to $1.8''$.  Bias frames and twilight flats were obtained every night.
Also, several Landolt (1992) standard fields were observed every night 
covering a range of airmasses.

Only one of the galaxies (Arp-loop) observed on 2000 February 9 was included
in Table~2, since the other galaxies observed during this night
were too heavily affected by clouds or no galaxy was visible in the field
(false detection).

\section{Data reduction}

The images were processed using the MIDAS
package developed at ESO.  After bias subtraction and flat-fielding, 
we removed cosmic ray hits using the {\sc filter/cosmic} task in MIDAS.  
Then images obtained in the same filter were co-added. 
The resulting V-band images are displayed in Fig.~1 (the galaxy KKH~23 
is given in the I filter).

Background stars were removed from the frames by fitting a
second-degree surface in circular pixel area. The sky background on the image
was approximated by a tilted plane, created from a 2-dimensional
polynomial, using the least-squares method ({\sc fit/flat\_sky}).
The uncertainties of the sky background determination were about 0.7\% of
the background level. The value of the surface brightness of the sky
is about 22 mag arcsec$^{-2}$ in the B band. Thus, the mean uncertainty
introduced
by the inaccuracy of the sky background determination is 
$\le 0\fm$15. The size of this uncertainty is primarily
caused by the background variations across the frame due to some difficulties
of flat-fielding.

The greater part of the sample is comprised of galaxies of extremely low
surface brightness. Many objects are characterized by an irregular, clumpy
structure. For this reason, we did not approximate the galaxies by
ellipses, but used circular apertures. The center of each galaxy was
determined interactively.

For the photometric calibration standard stars from
the list of Landolt (\cite{landolt}) were used. The observations
of the two nights were complicated by the appearance of clouds for 2--3
hours. However, for the rest of the time the conditions were quite photometric
judging from APO's $10\mu$ all-sky camera, which continuously monitors the
sky above APO.
To transform our instrumental magnitudes
to the standard Johnson-Cousins system, zero points and color coefficients
were determined. We used the extinction coefficient measured by Ted Wyder
({\tt http://www.apo.nmsu.edu/Instruments/Apozero-\\pts.html}). The uncertainties
of the transformation to the standard system are 0$\fm$05 in the B filter,
0$\fm$07 in the V filter, and 0$\fm$10 in the I filter.

\section{Total magnitudes and surface brightness profiles}

To measure total galaxy magnitudes in each band, integrated
photometry was performed in increasing circular apertures from a
pre-chosen center to the faint outskirts of the galaxies.
The total magnitude was then estimated as the asymptotic value of
the obtained radial growth curve.
The uncertainties of the total magnitude determination were $0\fm04$ in B,
$0\fm06$ in V,  and $0\fm09$ in I. The measurement results are
presented in Table~3. The first column contains the galaxy name, the 2nd
and 3rd columns give its total magnitude in the V filter and
the color index B--V (in the case of KKH~51 also V--I), without
correction for the Galactic extinction, the 4th and 5th columns show
the central surface brightness, the 6th column gives the absolute stellar
magnitude of the galaxies in the B filter corrected for the Galactic
extinction according to Schlegel et al.\ (\cite{schlegel}). To calculate
the distance, the radial velocity was used, and the Hubble constant was
assumed to be 70~km~s$^{-1}$~Mpc$^{-1}$. For the galaxies FM~1,
Arp-loop, and KKH~57 the distance modulus of M~81
of to 27.80 mag was taken, since their location makes it highly probable 
that the given objects belong to the M~81 group. The measurements of some
galaxies (marked with a colon) have low accuracy because of light clouds
or because of the presence of extremely bright stars in the image.
It does not seem possible to determine the parameters of UGC~7242
because of a bright star projected onto it.

For the investigation of dwarf galaxies azimuthally averaged surface
brightness profiles have been widely used
(Karachentseva et al. \cite{karachentseva96},
Papaderos et al. \cite{papaderos}, Bremnes et al. \cite{bremnes98}).
They allow one to improve the accuracy of surface photometry in galaxies
of low surface brightness and irregular structure.
Azimuthally averaged surface brightness profiles for our galaxies
were obtained by differentiating the galaxy growth curves with
respect to radius (Bremnes et al.\ \cite{bremnes98}). The resulting profiles
in the B (or I) and V colors are displayed in Fig.~2. Most of the galaxies
were measured up to the level of 28 to 29 mag arcsec$^{-2}$ in the B filter.
Mean uncertainties of the
measurements were estimated by intercomparison of individual
profiles for the same objects obtained from different frames
in the same passband.  They amount to about $0\fm05$ at the 23 mag 
arcsec$^{-2}$ isophotal level and about $0\fm3$ to $0\fm4$ at the 27 mag 
arcsec$^{-2}$ isophotal level in each of the filters.

By summing up the intrinsic errors indicated above, we derive the final
uncertainties of estimating total magnitudes~: 0$\fm$16 in the B filter,
0$\fm$18 in V, and 0$\fm$20 in I. The final uncertainties in estimating
of surface brightness depends on the SB level of the galaxy itself.
The errors at the level of 23 mag arcsec$^{-2}$ are 0$\fm$16
in the B filter, 0$\fm$17 in V, 0$\fm$19 in I. They rise to about 0$\fm$25 in B and V
at the level of 25--26 mag and amount to about 0$\fm$35 to 0$\fm$55 at
the level of 27--28 mag arcsec$^{-2}$. The uncertainties in the I filter are higher,
and reach to about 0$\fm$4 at the level of 26 mag arcsec$^{-2}$.

We found photometric data in the literature for only one
galaxy from our sample~: KK~78. The photometry was done by Hopp \& Schulte-Ladbeck
(\cite{hs91}). The agreement between the two sets of measurements
(B$_{our}$--B$_{HS91}$ = --0.08)
and ($\mu_B$(0)$_{our}$--$\mu_B$(0)$_{HS91}$ = 0.1) is satisfactory.

\section{Short description of the galaxies observed}

{\bf KKH~23}. This irregular dwarf galaxy of low surface brightness
has not been detected in H\,{\sc i} (Karachentsev et al. \cite{k2001a}).
The brightness profile in the V band is
well represented by an exponential. The photometry in the I band is
unreliable because of the residual background inhomogeneities.

{\bf KKH~35}. This is an object of extremely low surface brightness at a low
galactic latitude ($-4\fdg6$), showing no structural details. It is
utterly invisible in the I band and remains undetected in H\,{\sc i} (Karachentsev et al. \cite{k2001a}).
This is probably a faint planetary nebula in the Milky Way or Galactic cirrus.

{\bf KK~65}. According to Huchtmeier et al.(\cite{huchtmeier}),
the radial velocity of this irregular dwarf with respect to the Local Group
centroid is only V$_{LG}$ = +168 km~s$^{-1}$. This object is located
at a distance of 20$\arcmin$ from another irregular galaxy, UGC~3974,
which has nearly the same radial velocity (+161 km~s$^{-1}$).
The color index B--V of KK~65 increases slightly from center
towards periphery.

{\bf KKH~46}. The galaxy looks like a group of blue star-like condensations
embedded in a common envelope of low surface brightness. It has a corrected
radial velocity of +409 km~s$^{-1}$ (Karachentsev et al.\ \cite{k2001a}).
There are no other galaxies with close radial velocities within a projected
separation of $\sim$ 1.5 Mpc. KKH~46 can be treated as an example of a very isolated
dwarf system, where the ongoing star formation process is not initiated
by external tidal perturbations.

{\bf KKH~51}. Judging by the corrected radial velocity V$_{LG}$ =
438 km~s$^{-1}$ (Karachentsev et al.\ \cite{k2001a}) this galaxy is a
companion of the giant spiral NGC~2905, which has V$_{LG}$ = 443 km~s$^{-1}$.
The angular distance between them is 25$\arcmin$.

{\bf KKH~54 = UGC~5209}. The galaxy has a radial velocity of +479 km~s$^{-1}$
(Karachentsev et al. \cite{k2001a}). Together with UGC~5186 and
UGC~5272 this galaxy is likely to be a member of a very loose group
of irregular dwarf galaxies whose radial velocities lie in the interval
V$_{LG}$ = 460 -- 500 km~s$^{-1}$. The mean color index of the galaxy
increases from B--V = $0\fm48$ at the center to $0\fm60$ at the periphery, 
which may point to the presence of a more centrally concentrated young
population and a more extended old stellar population
in the galaxy.

{\bf FM~1}. This dwarf spheroidal galaxy of very low surface brightness
was recently discovered by Froebrich \& Meusinger (\cite{fm}) in
searching for new dwarf members of the M~81 group. Karachentsev
et al.\ (\cite{k2001b}) have resolved it into stars with the HST WFPC2
and estimated the distance to be 3.5 Mpc, which confirms the membership
of FM~1 in the M~81 group. Photometry of this galaxy
is affected by the presence of a bright star 2$\arcmin$ to the northwest from
its center.

{\bf KK~78 = UGC~5272b}. This blue compact galaxy is located 2$\arcmin$
southwest of another brighter dwarf system, UGC~5272, with V$_{LG}$ =
460 km~s$^{-1}$. From the spectroscopic measurements made by Makarov 
(private communication) its
radial velocity V$_{LG}$ = 466 km~s$^{-1}$, which may suggest that the
two galaxies are physically related.

{\bf A0952+69 = Arp-loop}. This is the brightest part of the
diffuse circular structure (Arp \cite{arp}) that embraces the northern
part of M~81. Efremov et al.\ (\cite{efremov}) have detected
quite a few bluish stars and condensations whose origin may be
due to the tidal interaction of M~81 with M~82 and
NGC~3077. In our CCD frame, two groups of faint objects can be 
distinguished: one in the east and one in the west with angular sizes of 
1$\farcm$0
and 0$\farcm$6, respectively. The photometric center was chosen to lie
at the eastern spot.
As one can see from the brightness profiles, the details of Arp-loop
have an extremely low surface brightness, which is fainter than 26 mag
arcsec$^{-2}$
in the V band. For this reason, the total magnitude $V_T$ and the integrated
color of the object cannot be determined with satisfactory accuracy.

{\bf KKH~57}. Just as FM~1 this galaxy is also a dSph member 
of the M~81 group. The HST observations made it possible to
resolve the galaxy into stars and yielded a distance of 3.7 Mpc
(Karachentsev et al. \cite{k2001b}).

{\bf KKSG~19}. This irregular galaxy of very low surface brightness
has a radial velocity V$_{LG}$ = +373 km~s$^{-1}$ (Huchtmeier et al.\
\cite{h2002}). The surface brightness profile shows a plateau in the central
part.  This galaxy shows a pronounced radial color gradient with an 
increasingly redder color towards the periphery.

{\bf KKH~67}. This galaxy of extremely low surface brightness is
tentatively classified as dIrr. Karachentsev et al.\ (\cite{k2001a}) have not
detected it in H\,{\sc i}. From the integrated color index (B--V)$_T = 1\fm05$
it looks more like a dSph than a dIrr.

{\bf KKH~70}. This is a compact irregular galaxy with blue condensations
on the northeastern side. The emission line with V$_{hel}$ = --154 km~s$^{-1}$
is likely to belong to local Galactic hydrogen (Karachentsev et al.\
 \cite{k2001a}).

{\bf KKSG~25}. This is a bluish galaxy of very low surface brightness.
Judging by its coordinates and radial velocity, V$_{LG}$ = +982 km~s$^{-1}$
(Huchtmeier et al.\ \cite{h2002}), it may be a member of the southern
extension of the Virgo cluster.

{\bf UGC~7242 = KKH~77}. An irregular dwarf galaxy with
a radial velocity of V$_{LG}$ = +213 km~s$^{-1}$ (Karachentsev et al. \cite{k2001a}). A very bright star
is projected onto its northern side, which strongly complicates
the photometry of the galaxy. Together with NGC~4236 and DDO~165
this galaxy forms the eastern spur of the M~81 group.

{\bf KKH~86}. This is a well isolated irregular galaxy with
a radial velocity of V$_{LG}$ = +205 km~s$^{-1}$ (Karachentsev et al.\ \cite{k2001a}).
Despite the small radial velocity value, it is not resolved into stars
and may belong to the periphery of the Virgo cluster.

\section{Discussion of results and conclusions}

It is well known that broad-band color indices are fundamental
for studying stellar populations in galaxies, and the color variations
along the radius reflect inhomogeneities of the stellar component.
About half of the galaxies under investigation demonstrate a minor increase
in the redness of the total color index B--V center to periphery. This is likely
to correspond to the increase in the average age of the stellar population
towards the edge of the galaxy. This property of many dwarf galaxies
was noticed earlier (Makarova \cite{makarova}). In particular, a more extended
old stellar population is a common property of dwarf
galaxies in the Local Group (Grebel \cite{grebel}). The absence of a noted
color gradient for half of the galaxies of our sample may be indicative
of a more homogeneous spatial distribution of the stellar populations of
different age in these dwarf galaxies, or of a small age and metallicity spread.

The median color index of the measured galaxies is 
$\langle B-V \rangle = 0\fm50\pm0\fm10$.
This color index is somewhat redder than in typical LSB galaxies, where
$\langle B-V \rangle = 0\fm45$ (McGaugh \& Bothun \cite{mb}; Vennik et al.
\cite{vennik}).

Figure~3 displays the relationship between central surface brightness of
the galaxies of the sample discussed and absolute stellar magnitude
in the B filter. All the values are corrected for Galactic extinction.
For comparison, similar relations for the spiral galaxies from the paper
by van der Kruit (\cite{kruit}), and also for the dwarf galaxies and
the galaxies of low surface brightness from the paper by Vennik et al.\
(\cite{vennik}) and for the dwarf galaxies from the article by
Makarova (\cite{makarova}) are plotted in the figure. As can be seen
from the figure the galaxies of the present study 
occupy a rather narrow range of
absolute B magnitudes, $\sim -10\fm0$ to $\sim -13\fm2$. 
Their central surface brightnesses
are distributed in the interval from 22 to 25.5 mag arcsec$^{-2}$. The mean
value of the central surface brightness in the B filter (corrected for
Galactic extinction) for our galaxies is 24.6~$\pm$~1.3 mag~arcsec$^{-2}$.
The nearby dwarf galaxies measured by Vennik et al.\ (\cite{vennik}) and
Makarova (\cite{makarova}) have on the average higher absolute magnitudes
and central surface brightnesses. The regions of nearby dwarf galaxies
and bright spirals are well-separated. Note that there
may be a weak correlation of the absolute stellar magnitude and the central
surface brightness for the galaxies indicated in the figure.
Such a correlation was noted earlier (Binggeli \& Cameron \cite{bc},
Vennik et al. \cite{vennik} and other authors). There is a noticeable
separation in the M -- $\mu$ diagram between galaxies of different
morphological types. It was suggested by Binggeli (\cite{b}) that the M -- $\mu$ diagram
for stellar systems might be the equivalent of the HR diagram for stars
(see Fig.~1 of his review).

McGaugh \& Bothun (\cite{mb}) believe that the distribution of galaxies
in the ($\mu_0$, M$_B$) plane may reflect the initial conditions of formation
of these objects. The luminosity corresponds approximately to the mass
enclosed in the density fluctuation from which the galaxy formed, while
the surface brightness corresponds to the density gradient in this fluctuation.

\begin{acknowledgements}
We are very grateful to D.I.~Makarov for kindly making available to us 
the results of the radial velocity measurement of KK~78.
IDK and EKG would like to thank the staff at Apache Point Observatory 
for their support during the on-site and remote observations. 
The work of LNM was supported by INTAS grant YSF 2001/1-0129.
IDK and EKG acknowledge support through the
Henri Chr\'etien International Research Grant administered by the American
Astronomical Society, which made it possible to obtain the data presented 
here.  EKG was supported by NASA through grant
HF-01108.01-98A from the Space Telescope Science Institute during part of
this work.
\end{acknowledgements}

{}

\newpage

\begin{table*}[h]
\caption{General parameters of the studied galaxies}
\begin{tabular}{lccccccc}
\hline
Object &$\alpha_{2000}$ &$\delta_{2000}$                   & A$_B$& A$_V$&Type$^\ast$ &$a$&$V_{LG}$ \\
       &                &                                  & mag  & mag  &           & arcmin & km s$^{-1}$ \\  \hline
KKH23  &$03^h42^m53.^s5$&$38^\circ37^{\arcmin}00^{\arcsec}$&1.17 &0.89 &Irr, L      & 1.1        &      \\
KKH35  &05 57 07.8      &15 25 28                          &1.71 &1.30 &Irr, EL     & 1.3        &      \\
KK65   &07 42 32.0      &16 33 39                          &0.14 &0.11 &Irr, H      & 0.9        &168   \\
KKH46  &09 08 36.6      &05 17 32                          &0.20 &0.15 &Irr, L      & 0.6        &409   \\
KKH51  &09 30 12.9      &19 59 30                          &0.19 &0.15 &Irr, L      & 0.7        &438   \\
KKH54  &09 45 03.9      &32 14 17                          &0.08 &0.06 &Irr, L      & 0.9        &479   \\
FM1    &09 45 10.0      &68 45 54                          &0.34 &0.26 &Sph, VL     & 1.0        &      \\
KK78   &09 50 19.6      &31 27 24                          &0.09 &0.07 &Irr, H      & 0.5        &466   \\
Arp-loop&09 57 29.0      &69 16 20                          &0.37 &0.28 &Irr, EL    & 1.8        &      \\
KKH57  &10 00 16.0      &63 11 06                          &0.10 &0.08 &Sph, VL     & 0.6        &      \\
KKSG19  &10 24 28.3      &-12 25 57                         &0.30 &0.23 &Irr, VL    & 0.7        &373   \\
KKH67  &11 23 03.5      &21 19 37                          &0.10 &0.08 &Irr, EL     & 1.2        &      \\
KKH70&11 39 26.9&60 10 18                          &0.06 &0.05 &Irr, H      & 0.9        &      \\
KKSG25  &11 45 17.8      &-17 16 26                         &0.17 &0.13 &Irr, VL     & 1.2        &982   \\
UGC 7242 &12 14 07.3 & 66 05 32                            &0.08 &0.06 &            & 1.5        &213   \\
KKH86  &13 54 33.5      &04 14 35                          &0.12 &0.09 &Irr, L      & 0.7        &205  \\
\hline
\multicolumn{7}{l}{$\ast$ estimated by I.D.Karachentsev}\\
\multicolumn{7}{l}{Irr - irregular}\\
\multicolumn{7}{l}{Sph - spheroidal}\\
\multicolumn{7}{l}{EL - extremely low surface brightness}\\
\multicolumn{7}{l}{VL - very low surface brightness}\\
\multicolumn{7}{l}{L - low surface brightness}\\
\multicolumn{7}{l}{H - high surface brightness}\\
\end{tabular}
\end{table*}

\begin{table*}[h]
\caption{Observational log of the nearby dwarf galaxies}
\begin{tabular}{lcccc}
\hline
 Object & Date   &Filter & Exposure& Airmass \\
	&         &       & sec       &          \\ \hline
 KKH23  & 03022000& V     & 300       & 1.146\\
	&         & I     & 2$\times$300 & 1.168\\
 KKH35  & 03022000& V     & 2$\times$300 & 1.078\\
	&         & I     & 2$\times$300 & 1.077\\
 KK65   & 03022000& B     & 300       & 1.392 \\
	&         & V     & 300       & 1.440 \\
 KKH46  & 03022000& B     & 2$\times$300 & 1.204\\
	&         & V     & 2$\times$300 & 1.182\\
 KKH51  & 03022000& B     & 300       & 1.250\\
	&         & V     & 2$\times$300 & 1.309\\
	&         & I     & 2$\times$300 & 1.307\\
 KKH54  & 0302200 & B     & 2$\times$300 & 1.185\\
	&         & V     & 2$\times$300 & 1.184\\
 FM1    & 03022000& B     & 2$\times$300 & 1.396 \\
	&         & V     & 2$\times$300 & 1.418 \\
 KK78   & 03022000& B     & 300       & 1.070\\
	&         & V     & 300       & 1.079\\
 Arp-loop& 09022000& B     & 300       & 1.334\\
	&         & V     & 4$\times$300 & 1.371\\
 KKH57  & 03022000& B     & 2$\times$300 & 1.456\\
	&         & V     & 3$\times$300 & 1.398\\
 KKS19  & 03022000& B     & 2$\times$300 & 1.460\\
	&         & V     & 2$\times$300 & 1.483\\
 KKH67  & 03022000& B     & 2$\times$300 & 1.298\\
	&         & V     & 2$\times$300 & 1.294\\
 KKH70  & 03022000& B     & 2$\times$300 & 1.131\\
	&         & V     & 2$\times$300 & 1.136\\
 KKS25  & 03022000& B     & 2$\times$300 & 1.561\\
	&         & V     & 2$\times$300 & 1.558\\
UGC 7242& 03022000& B     & 300          & 1.199\\
	&         & V     & 300          & 1.198\\
 KKH86  & 03022000& B     & 2$\times$300 & 1.158\\
	&         & V     & 2$\times$300 & 1.170\\
\hline
\end{tabular}
\end{table*}

\begin{table*}[h]
\caption{Measurement results for the galaxies of the sample}
\medskip
\begin{tabular}{lcccccc}
\hline
Object & $V_T$ &$(B-V)_T$&$\mu_B$(0) & M$_B$ \\
       &       &$(V-I)_T$&           & \\ \hline
KKH23  &16.00 :&         &           & \\
       &       & 0.93 :  &           & \\
KKH35  &15.87  &         &           & \\
KK65   &14.72  &0.54     & 22.5      & --11.78\\
KKH46  &16.67  &0.38     & 24.3      & --11.98\\
KKH51  &16.48  &0.54     & 24.5      & --12.15\\
       &       &0.77     &           & \\
KKH54  &15.50  &0.56     & 23.8      & --13.20\\
FM1    &16.62  &0.88     & 25.4      & --10.64\\
KK78   &17.28  &0.40     & 22.6      & --10.21\\
Arp-loop&16.08 :&0.72 :  & 26.9      & --11.41\\
KKH57  &17.12  &0.75     & 25.3      & --9.98\\
KKSG19  &16.19  &0.60     & 25.4      & --12.14\\
KKH67  &16.82 :&1.05 :   & 25.7      & \\
KKH70  &15.87  &0.60     & 23.5      & \\
KKSG25  &17.27  &0.54     & 25.5      & --13.10\\
KKH86  &16.21  &0.67     & 24.3      & --10.51\\
\hline
\end{tabular}
\end{table*}

\clearpage

\begin{figure*}[h]
\caption{The images of the sample galaxies, which were obtained
at the 3.5-m telescope of Apache Point Observatory.
The field size of the images is 4$\farcm$8$\times$4$\farcm$8.
For all the images North is top and East is left.}
\end{figure*}

\begin{figure*}[p]
\psfig{figure=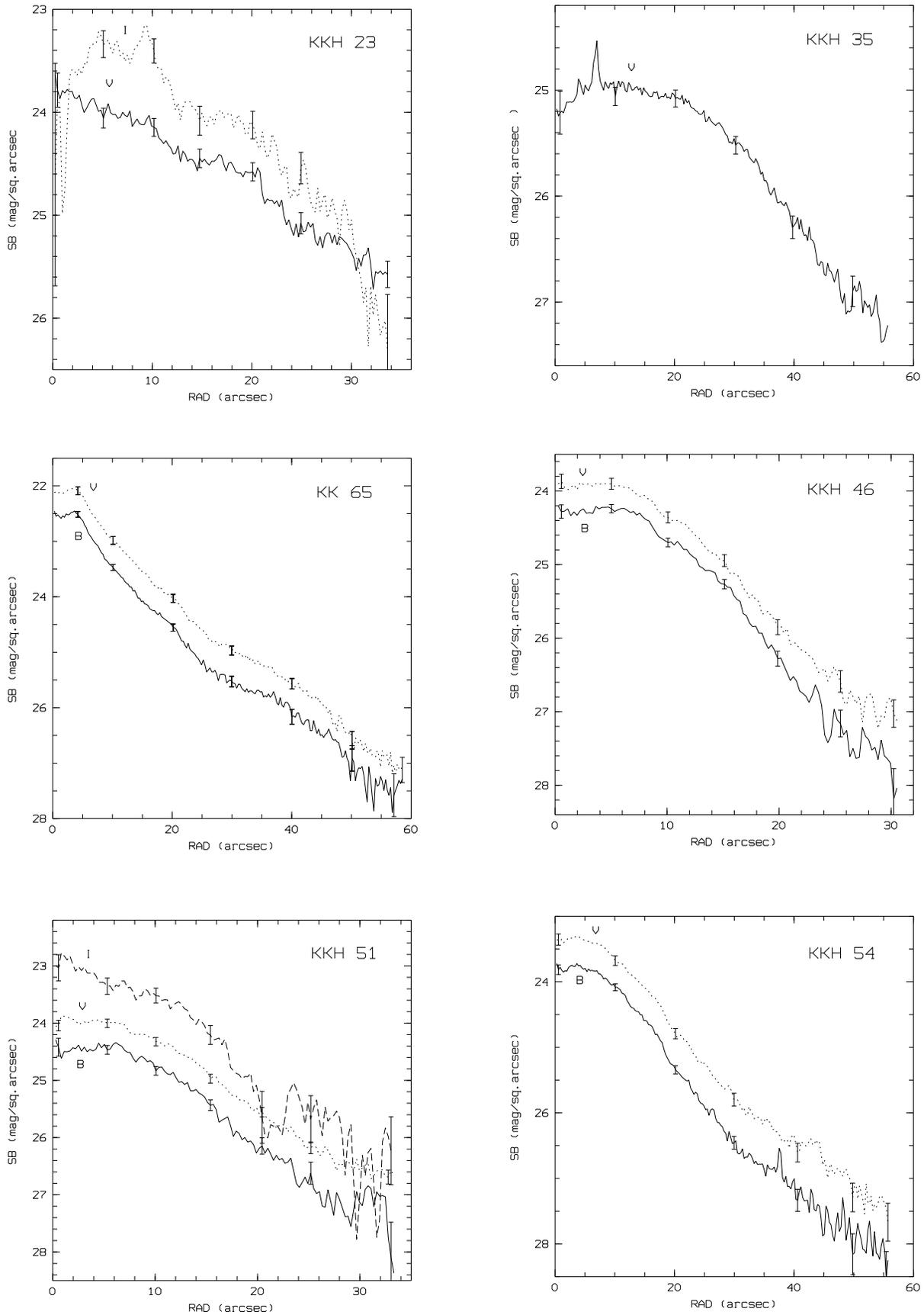,width=16cm}
\caption{The azimuthally averaged surface brightness
profiles for the nearby dwarf galaxies.}
\end{figure*}

\setcounter{figure}{1}
\begin{figure*}[p]
\psfig{figure=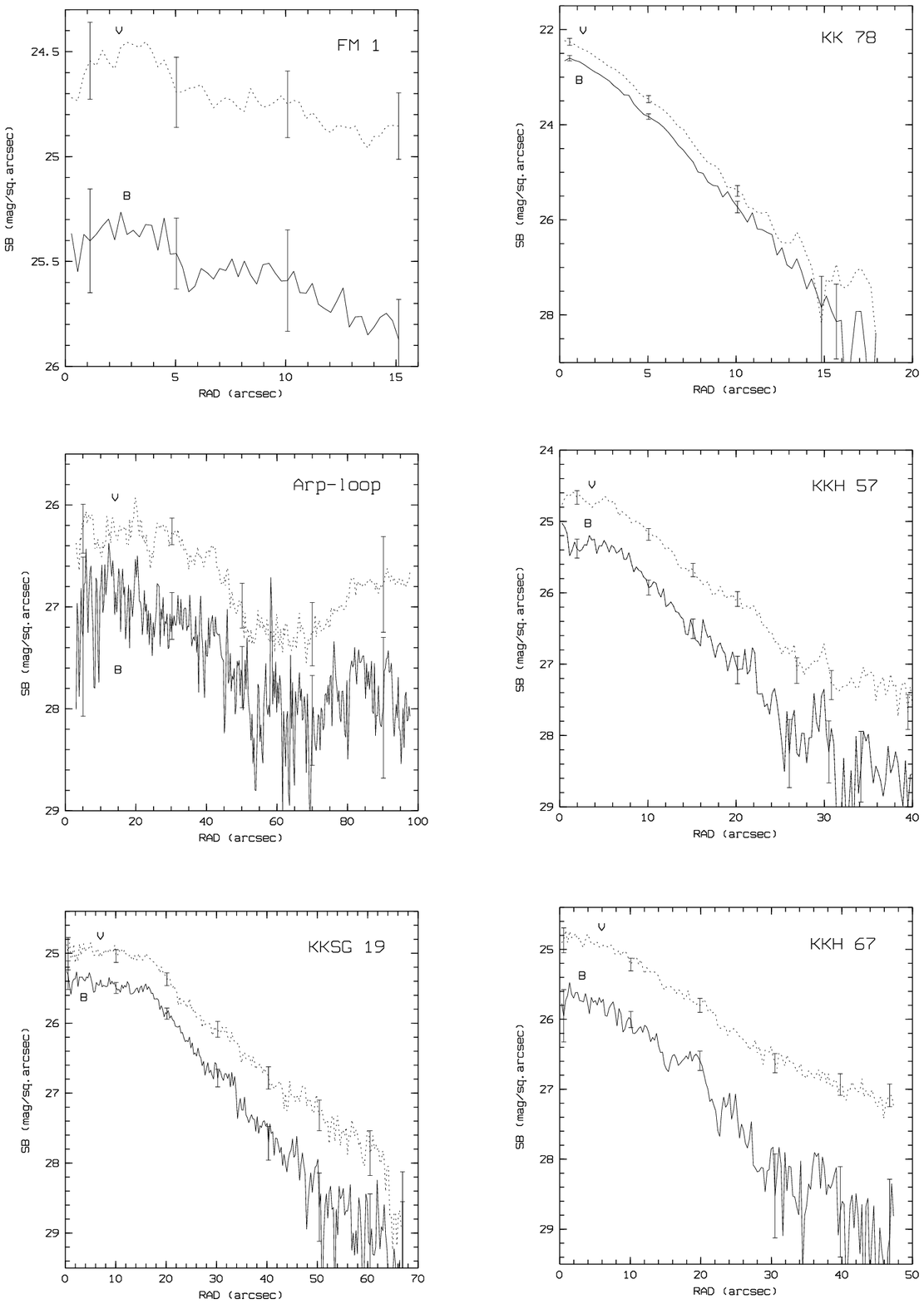,width=16cm}
\caption{continued.}
\end{figure*}

\setcounter{figure}{1}
\begin{figure*}[p]
\psfig{figure=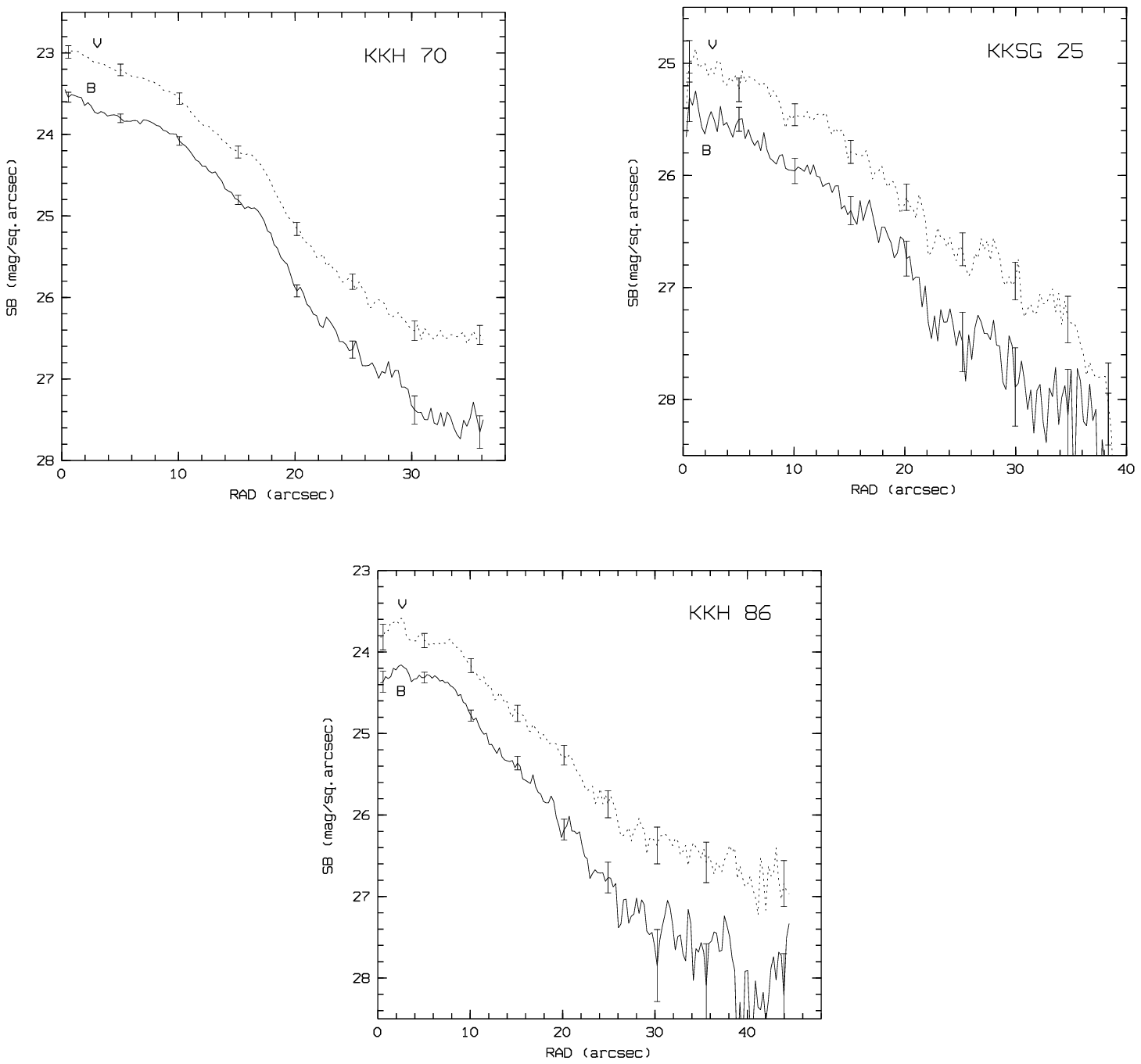,width=16cm}
\caption{continued.}
\end{figure*}

\setcounter{figure}{2}
\begin{figure*}[h]
\psfig{figure=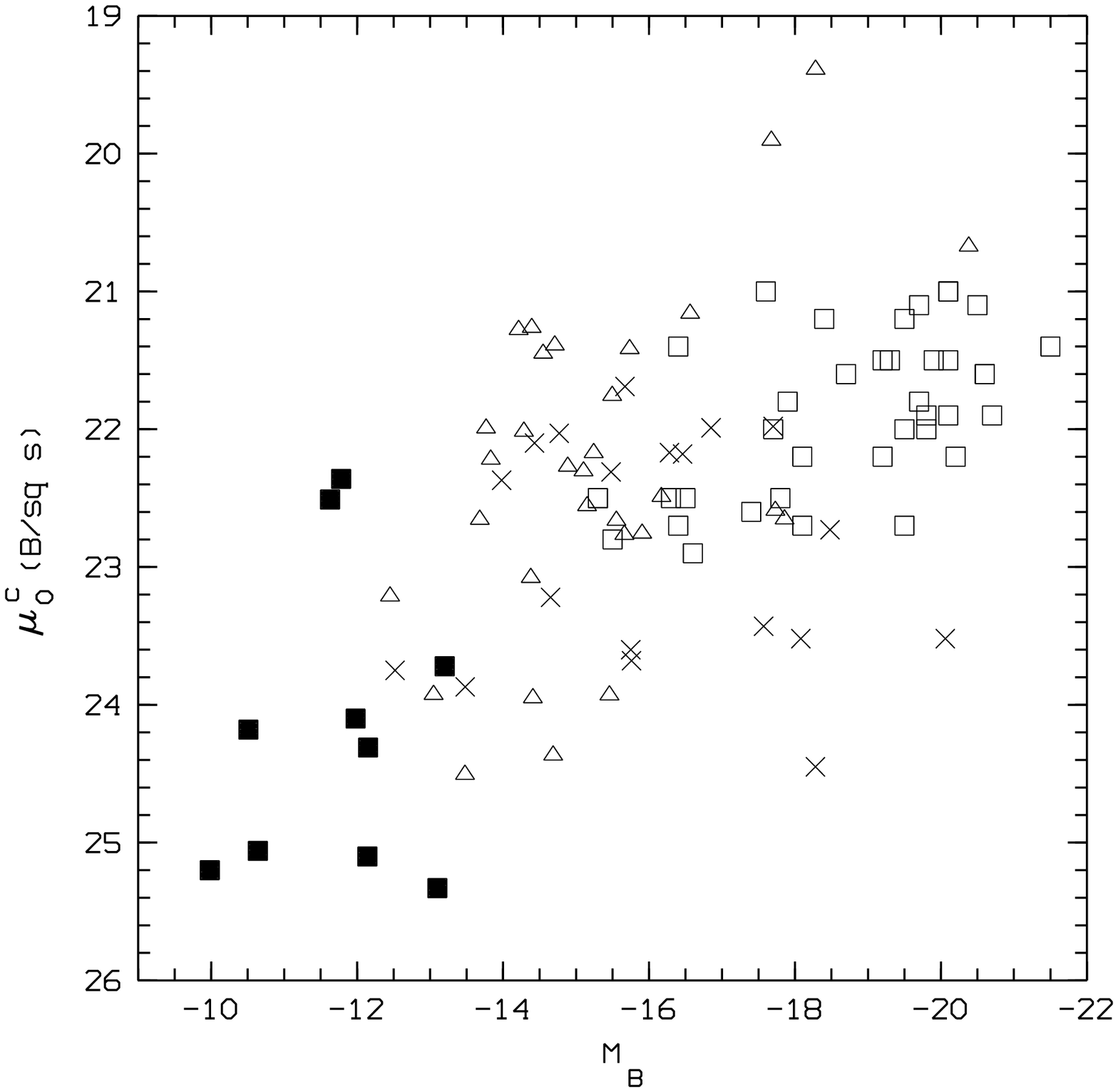,width=18cm,angle=-90}
\caption{The central surface brightness in B band versus
the absolute B-magnitude. The filled squares are our measured galaxies,
the triangles are the galaxies from the paper by Makarova (1999),
crosses are the low surface brightness galaxies from the article by
Vennik et al. (1996), and open squares are the spiral galaxies
from the work of van der Kruit (1987).}
\end{figure*}

\end{document}